\documentstyle[11pt]{article}
\begin{document}
\begin{flushright}SJSU/TP-95-13\\August 1995\end{flushright}
\vspace{1.7in}
\begin{center}\Large{\bf Can Future Events Influence the Present?}\\
\vspace{1cm}
\normalsize\ J. Finkelstein\footnote[1]{
        Participating Guest, Lawrence Berkeley National Laboratory\\
        \hspace*{\parindent}\hspace*{1em}
        e-mail: FINKEL@theor3.lbl.gov}\\
        Department of Physics\\
        San Jos\'{e} State University\\San Jos\'{e}, CA 95192, U.S.A
\end{center}
\begin{abstract}
Widom, Srivastava, and Sassaroli [Phys. Lett. A {\bf 203}, 255 (1995)]
have published a calculation which purports to show that
``future events can affect present events''.  In this note an error
in their calculation is identified.
\end{abstract}
\newpage
Widom, Srivastava, and Sassaroli [ref. 1, hereafter called WSS]
have recently published
a claim that quantum theory implies that
``future events can affect present events''.  WSS analyze an
experiment in which  photons are incident on a partially reflecting
plate, from which they are either reflected toward a detector $D_1$
or transmitted toward a detector $D_2$.  According to WSS, the
counting rate at $D_1$ can depend on the time at which the photons
reach $D_2$, even in cases in which the counts at $D_1$ are recorded
earlier than the time at which the photons reach
$D_2$.  This would seem to imply that
one experimenter, by varying the position of $D_2$, could send a
message backward in time to another experimenter located at $D_1$.

In this note, without addressing the plausibility of this claim,
we will point out an error in the calculation done by WSS.
We first briefly summarize the argument given by WSS. They
consider photons initially
heading toward the plate in wave packets called
$|\alpha\rangle$ and $|\beta\rangle$.
Upon hitting the plate, state $|\alpha\rangle$ evolves into two packets:
$|\alpha_{1}\rangle$ which is heading for $D_1$
and $|\alpha_{2}\rangle$ which is heading
for $D_2$.  Likewise, state $|\beta\rangle$ evolves into
$|\beta_{1}\rangle$ and $|\beta_{2}\rangle$.
It is essential to the argument of WSS that
the photon states $|\alpha\rangle$ and $|\beta\rangle$
not be orthogonal.  Defining $\epsilon$ to be the overlap of the
packets, we can write (as in eq.\ 3 of WSS)
\begin{equation}
    \epsilon = \langle \alpha_{1} |\beta_{1}\rangle
    + \langle \alpha_{2} |\beta_{2}\rangle.
\end{equation}
WSS argue that the counting rate at $D_1$ can depend on
(among other things) the value of $\epsilon$, and that the
value of $\epsilon$ can depend on the time of arrival of the
photons at $D_2$ (as in eq.\ 8 of WSS).
This would imply that a variation in the
time of arrival at $D_2$ (e.g., a variation in the position
of $D_2$) would affect the counting rate at $D_1$.
To calculate $\epsilon$, WSS approximate the wave packets by
plane waves, of frequencies $\omega_{\alpha}$ and
$\omega_{\beta}$.  This introduces factors of
$\exp [i(\omega_{\alpha}-\omega_{\beta})t]$
in both terms on the RHS of eq.\ 1 above; then evaluating
$\epsilon$ at the times at which the photons arrive at the
detectors leads to the claimed result. Fig.\ 3
of WSS illustrates the calculated dependence of the counting
rate at $D_1$ upon the arrival time at $D_2$.

Of course, as WSS
realize, the packets cannot really be plane waves of distinct
frequencies; the condition $\epsilon \neq 0$ requires that the
frequency spectra of the packets $|\alpha\rangle$ and
$|\beta\rangle$ must overlap. Fortunately, there is no need to
use a plane wave approximation, or indeed any
approximation, for the time development of the packets.
It is exactly true that the value of
$\epsilon$, as well as the value of each term on the
RHS of our eq.\ 1, is independent of the time.  For example,
letting $U$ represent the time-development operator
(which for a free wave packet is just a translation at speed $c$),
we can write $|\alpha_{2}(t)\rangle
= U(t,t_{0})|\alpha_{2}(t_{0})\rangle $ and
$|\beta_{2}(t)\rangle = U(t,t_{0})|\beta_{2}(t_{0})\rangle $, and so
\begin{eqnarray}
     \langle\alpha_{2}(t)|\beta_{2}(t)\rangle & =
     & \langle\alpha_{2}(t_{0})| U^{\dagger}(t,t_{0})
              U(t,t_{0})|\beta_{2}(t_{0})\rangle \nonumber \\
          & = & \langle\alpha_{2}(t_{0})|\beta_{2}(t_{0})\rangle;
\end{eqnarray}
thus $\langle\alpha_{2}|\beta_{2}\rangle$ is independent of time.
In fact, to get the value of $\epsilon$, one need only consider that
\begin{equation}
      \epsilon = \langle\alpha|\beta\rangle
\end{equation}
where $|\alpha\rangle$ and $|\beta\rangle$
are the packets before they have hit the plate.

Either from our eq.\ 3 (evaluated before the photons hit the
plate) or our eq.\ 1 (evaluated right after they hit the plate)
one can see that the value of $\epsilon$ does not depend on the
time at which the photons reach $D_2$. This means that eq.\ 8
of WSS, which does exhibit such a spurious dependence,
cannot be correct.  Once the initial packets are chosen,
the value of $\epsilon$ is fixed; it cannot be varied by changing
the time of arrival at $D_2$.  Of course one could imagine choosing
to take different packets, corresponding to different
positions of $D_2$, but then the influence on the counting rate
at $D_1$ would come from the earlier choice of the packets,
not from the (possibly later) choice of the position of $D_2$.

Therefore the calculation presented by WSS does not support
their conclusion that future events can affect present events.
Of course, this does not rule out the possibility that their
conclusion might nevertheless be correct.  If that should turn
out to be the case, I will not have to suffer any embarrassment
for having written this note; if at some point in the future I
learn that the conclusion of WSS was in fact correct, I will merely
have to use a mechanism similar to the one they suggest to make
sure that this note never did get written!

\vspace{1cm}
Acknowledgement:
I would like to acknowledge the hospitality of the
Lawrence Berkeley National Laboratory.


\begin{thebibliography}{9}
\bibitem{1} A. Widom, Y.N. Srivastava, and E. Sassaroli,
            Phys. Lett. A {\bf 203}, 255 (1995).
\end{thebibliography}
\end{document}